\documentclass[aps,prl,amsmath,twocolumn,amssymb,superscriptaddress]{revtex4}
\pdfoutput=1
\usepackage{graphicx}
\usepackage{epstopdf}
\usepackage{color}
\usepackage{cases,array}

\begin{document}
\title{Non-Adlerian synchronization of dipolar coupled vortex Spin-Torque Nano-Oscillators}

\author{A.D. Belanovsky}
\affiliation{A. M. Prokhorov General Physics Institute, RAS, Vavilova, 38, 119991 Moscow, Russia and Moscow Institute of Physics and Technology, Institutskiy per. 9, 141700 Dolgoprudny, Russia}

\author{N. Locatelli}
\affiliation{Unit\'e Mixte de Physique CNRS/Thales, 1 ave A. Fresnel, 91767 Palaiseau, and Univ Paris-Sud, 91405 Orsay, France}

\author{P.N. Skirdkov}
\affiliation{A. M. Prokhorov General Physics Institute, RAS, Vavilova, 38, 119991 Moscow, Russia and Moscow Institute of Physics and Technology, Institutskiy per. 9, 141700 Dolgoprudny, Russia}

\author{F. Abreu Araujo}
\affiliation{Universit\'e catholique de Louvain, 1 Place de l'Universit\'e, 1348 Louvain-la-Neuve, Belgium}

\author{K.A. Zvezdin}
\affiliation{A. M. Prokhorov General Physics Institute, RAS, Vavilova, 38, 119991 Moscow, Russia and Moscow Institute of Physics and Technology, Institutskiy per. 9, 141700 Dolgoprudny, Russia}
\affiliation{Istituto P.M. srl, via Grassi, 4, 10138, Torino, Italy}

\author{J. Grollier}
\affiliation{Unit\'e Mixte de Physique CNRS/Thales, 1 ave A. Fresnel, 91767 Palaiseau, and Univ Paris-Sud, 91405 Orsay, France}

\author{V. Cros}
\affiliation{Unit\'e Mixte de Physique CNRS/Thales, 1 ave A. Fresnel, 91767 Palaiseau, and Univ Paris-Sud, 91405 Orsay, France}

\author{A.K. Zvezdin}
\affiliation{A. M. Prokhorov General Physics Institute, RAS, Vavilova, 38, 119991 Moscow, Russia and Moscow Institute of Physics and Technology, Institutskiy per. 9, 141700 Dolgoprudny, Russia}

\begin{abstract}
We investigate analytically and numerically the synchronization dynamics of dipolarly coupled vortex based Spin-Torque Nano Oscillators (STNO) with different pillar diameters. 
We identify the critical interpillar distances on which synchronization occurs as a function of their diameter mismatch.
We obtain numerically a phase diagram showing the transition between unsynchronized and synchronized states and compare it to analytical predictions we make using Thiele approach. Our study demonstrates that for relatively small diameters differences the synchronization dynamics can be described qualitatively using Adler equation. However when the diameters difference increases significantly, the system becomes strongly non-Adlerian.

\end{abstract}

\maketitle

The study of synchronization process is an important problem of nonlinear science, not only because of the wide range of  applications in physics, biology, chemistry, and even in social systems, but also because of a numerous fundamental challenges in understanding of  collective dynamics in large ensembles. Recently a great attention has been drawn to the studies of the phase locking in the arrays of Spin-Torque Nano-Oscillators (STNO)\cite{Georges2008-2,Zhou2009}. STNO benefit from the spin-transfer phenomenon \cite{Slonczewski1999,Berger1996,Kiselev2003} to generate the precession of magnetization in a free magnetic layer, and magnetoresistance effects to get the corresponding voltage signal. These generators are nanoscaled, easily compatible with CMOS cycle, and can be easily tuned by dc current and/or external magnetic field.

Here we consider STNOs with free layers being in a vortex state, where the gyrotropic motion of vortex core can be excited by the injection of a dc current through a STNO stack, even without any applied external magnetic field \cite{Flavio2012}. Using both the gyrotropic vortex core motion and the tunnel magnetoresistance (TMR) it is possible to obtain coherent and high power output microwave signals \cite{Dussaux2010}. Interest for synchronization of STNOs has arisen initially from the need to increase further the coherence of the magnetic oscillations to match the requirements of telecommunication applications, but is increasingly considered as a strong candidate for coherent multiple spin-wave emissions \cite{Demidov2010} as well as for networks of oscillators for associative memory applications\cite{Csaba2012,Horvath2012}.

\begin{figure}[ht!]
\centerline{\includegraphics[width=8.5cm]{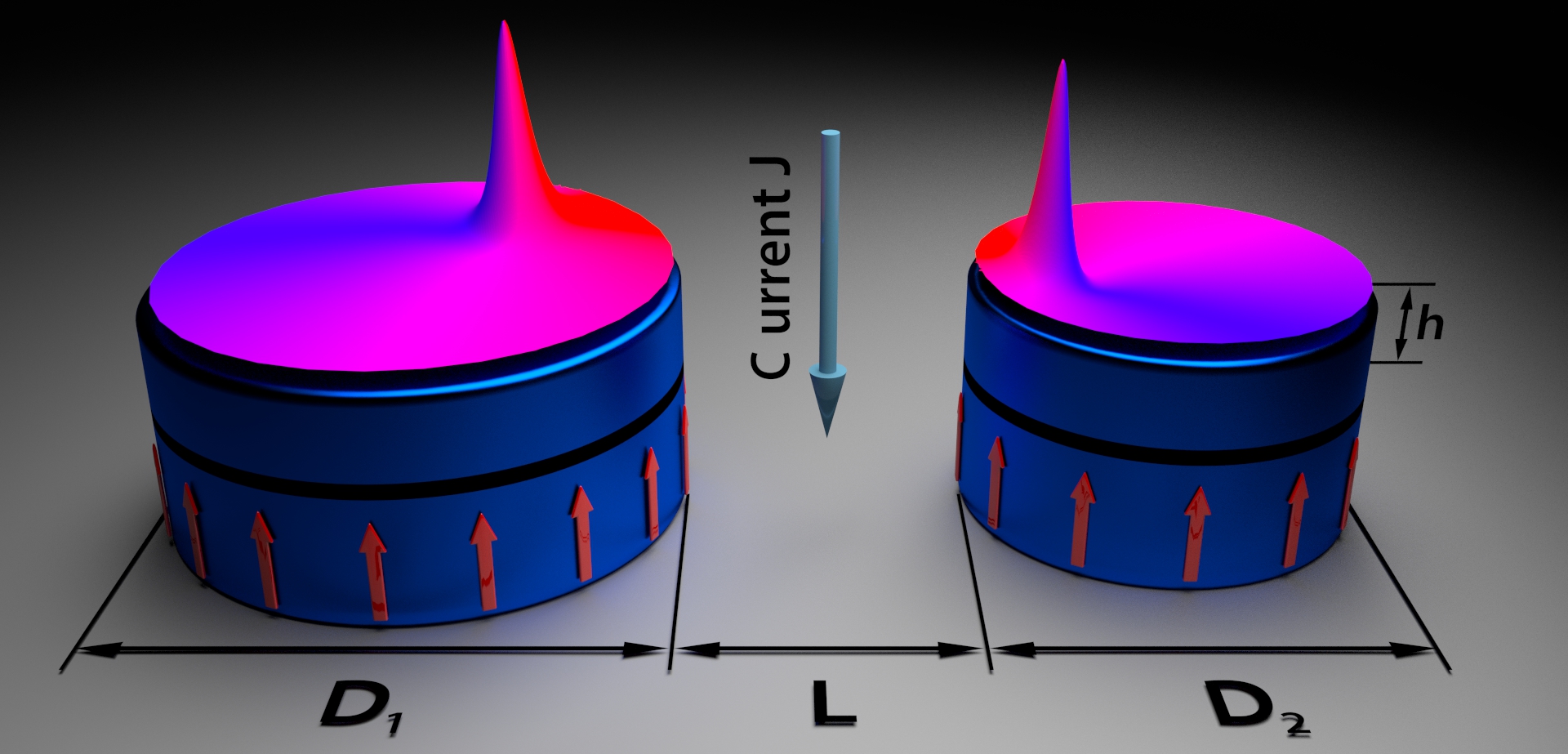}}
\caption{\footnotesize{(Color online) Schematic representation of the studied system. There are two STNOs with diameters $D_{1,2} = D_0 \pm \Delta D/2$, where $D_0 = 200$ nm and $\Delta D/D_0~(\Delta D = D_1 - D_2)$ is a diameters detuning which is not more than 15\%.}}
\label{fig:fig1}
\end{figure}

\begin{figure*}[ht!]
\centerline{\includegraphics[width=17cm]{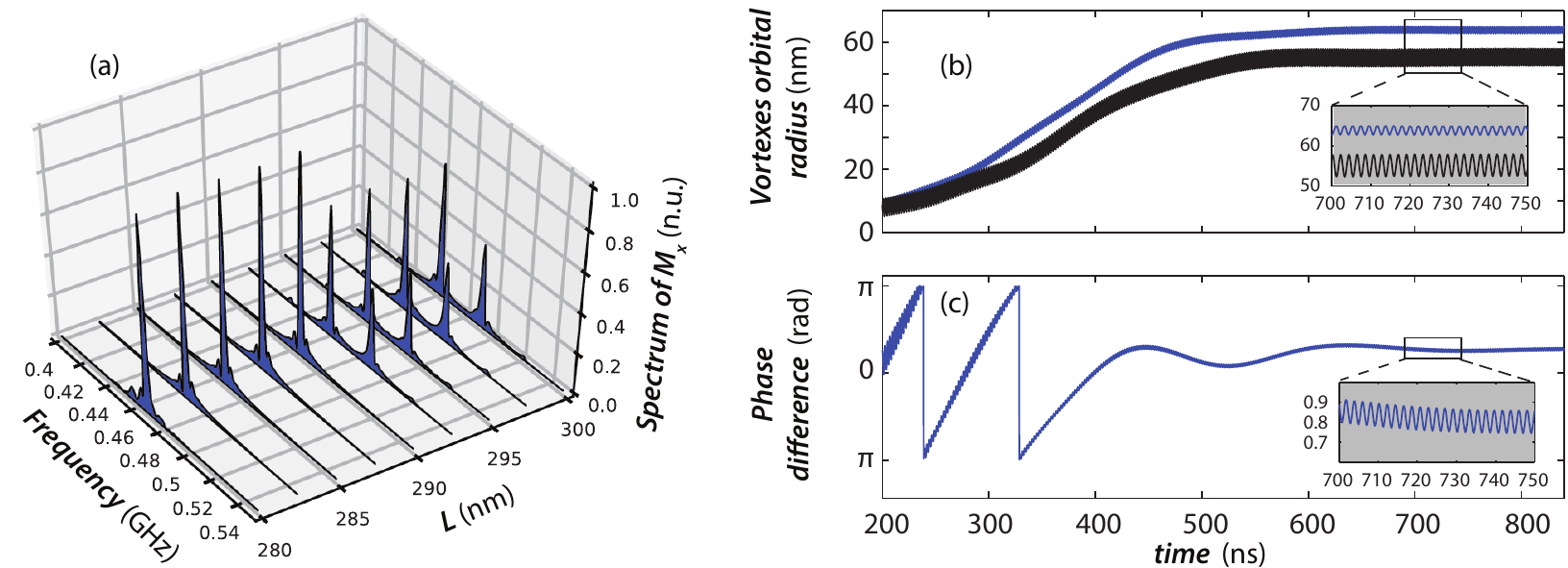}}
\caption{\footnotesize{(Color online) a) Power spectrum density of $x$-component of total magnetization as a function of distance between disks edges for $\Delta D / D_0 = 10 \%$, showing the transition between unsynchronized and synchronized states b) Core positions $X_{1,2}$ vs time for $\Delta D / D_0 = 2.5 \%,~L = 600$ nm. c) Phase difference as a function of time, ($\Delta D / D_0 = 2.5 \%,~L = 600$ nm).}}
\label{fig:fig2}
\end{figure*}

There are several remarkable experimental studies of STNOs synchronization achieved by various physical mechanisms: through electrical connection in series of STNOs \cite{Tiberkevich2009, Georges2008-1,Grollier2006}, by spin wave propagation \cite{Chen2009, Berkov2013} and by antivortices \cite{Zaspel2013,Ruotolo2009}. Another important mechanism to synchronize STNOs is the magnetostatic coupling \cite{Shibata2003, Vogel2010, Awad2010, Barman2010, Jung2010}. Theoretical description of the synchronization dynamics of vortex STNO is more complicated task than the well known oscillating systems such as coupled Van der Pol oscillators or rotators (Josephson junction, rotating pendula) which have fixed orbit radius i.e. one degree of freedom, and thus can be described by Adler equation \cite{Pikovsky}. 

In this paper, we investigate the feasibility of synchronization through dipolar interaction focusing on the case of spin transfer vortex oscillators (STVOs). The vortex oscillator is  a model system of particular interest since it only couples to neighbors when out of equilibrium due to  the almost zero mean magnetization at equilibrium. The orbit radius in STVOs may change in time at the limit cycle, making the phase spaces to be multidimensional. Thus the second aim is to understand the workability of simplified analytical descriptions based on Thiele equation for the synchronization dynamics studies.

We consider a system made of two nanopillars (see Fig. \ref{fig:fig1}), each of them being composed by a free magnetic layer, a non magnetic spacer, and a fixed polarizer which generates a perpendicular spin polarization $p_z$. Free layers in both STVOs are $h = 10$ nm thick Ni$_{81}$Fe$_{19}$\footnote{The magnetic parameters of the free layers are: the magnetization $M_s$ = 800 emu/cm$^3$, the exchange constant $A = 1.3\times 10^{-6}$ erg/cm and the damping parameter $\alpha = 0.01$. The micromagnetic simulations have been performed by numerical integration of the Landau-Lifshitz-Gilbert equation using our micromagnetic code SpinPM based on the forth order Runge-Kutta method with an adaptive time-step control for the time integration and a mesh size $2.5 \times 2.5~\text{nm}^2$.}. The initial magnetic configuration are two centred vortices with same core polarities and same chiralities. The polarizing layers are not included in our consideration because these layers, being uniformly magnetized in $z$ direction, have almost no influence on the vortices motion.

In order to put STVOs in a steady oscillations regime, equal current densities $J$ of $7\times10^6$ A/cm$^2$ with the identical spin polarization $p_z = 0.2$ which corresponds to spin-torque amplitude $a_J = 5$ Oe are injected through both STVOs. The difference in initial frequencies of the two oscillators $f^{i}_{1}$ and $f^{i}_{2}$ arise because of the deliberate pillar diameters difference $\Delta D = D_1 - D_2$ around a mean diameter $D_0=200$ nm. Such system is close to real device situation, when STVOs are not identical because of lithography process accuracy.

\begin{table}[ht!]
\centering
\begin{ruledtabular}
\begin{tabular}{>{\centering\arraybackslash}m{1cm}|>{\centering\arraybackslash}m{1cm}|>{\centering\arraybackslash}m{1.2cm}|>{\centering\arraybackslash}m{1cm}|>{\centering\arraybackslash}m{1cm}|>{\centering\arraybackslash}m{1cm}|>{\centering\arraybackslash}m{1cm}}
$D_1$ (nm) & $D_2$ (nm) & $\Delta D / D_0$ $(\%)$ & $L_{cr}$ (nm) &$f_{1}^i$ (MHz) & $f_{2}^i$ (MHz)& $f_{12}$ (MHz)\\
\hline
202.5 & 197.5 & 2.5 & 607.5 & 473.1 & 480.6 & 477.8\\
205 & 195 & 5 & 495 & 469.4 & 484.7 & 484.7 \\
207.5 & 192.5 & 7.5 & 397.5 & 465.4 & 488.9 & 469.8\\
210 & 190 & 10 & 287.5 & 462.2 & 493.7 & 468.1\\
212.5 & 187.5 & 12.5 & 150 & 458.0 & 498.0 & 463.5\\
215 & 185 & 15 & 90 & 457.0 & 501.9 & 456.5
\end{tabular}
\end{ruledtabular}
\caption{\footnotesize{Diameters difference and critical distance $L_{cr}$ where synchronization still exists. Frequencies $f_1^i$, $f_2^i$ correspond to the frequencies of isolated STVOs and $f_{12}$ is the common frequency after synchronization.}}
\label{table1}
\end{table}

In this work the series of micromagnetic simulations have been performed for diameters corresponding to differences between 2.5$\%$ and 15$\%$. For each case, we varied interpillar distance with a step of 2.5 nm in order to determine the critical distance at which STVOs are still synchronized. The results of these simulations are summarized in table \ref{table1}. On figure \ref{fig:fig2} a) we present the evolution of the spectrum  of the $M_{x}$ component of the magnetization with decreasing distance between disks edges for the case $D_1 = 210$ nm, $D_2 = 190$ nm . As illustrated, when $L<L_{cr}$ the magnetization dynamics spectrum of the system eventually shows a transition between two independent peaks and a single one at frequency $f_{12}$, demonstrating synchronization achievement. The synchronized frequency $f_{12}$ stands between the frequencies of the isolated auto-oscillators, with a value closer to the frequency of the bigger oscillator. To illustrate the synchronization process, the core dynamics in the case $D_1 = 202.5$ nm, $D_2 = 197.5$ nm, and $L=600$ nm ($L<L_{cr}$) are shown in Fig. \ref{fig:fig2} b) (core orbit radii $X_1$, $X_2$) and c) (phase difference). As the vortices orbits increase towards their steady state values, the dipolar interaction between the two vortices increases. Once the interaction energy becomes strong enough, a phase locking occurs, then leading to convergence of the phase difference to a constant value,  meaning that the oscillators are synchronized.

As can been seen in Fig. \ref{fig:fig2} b) and c) inserts, high frequency oscillations of the orbit radii and the phase difference can be observed, and even remains after the synchronization is achieved. They are associated to high frequency forces acting on the cores in their gyrotropic rotating frame induced by the dipolar interaction. Notably, their action is averaged over low-frequency synchronization dynamics, and will be further neglected in the development of our models.

Table \ref{table1} presents the evolution of the frequency of synchronized oscillators $f_{12}$ with diameters difference, relatively to the isolated frequencies $f_{1}^i$ and $f_{2}^i$. For diameters difference below 10\%, we found that critical distances are greater than 250 nm, demonstrating the high efficiency of dipolar coupling for vortex oscillators synchronization. Mostly, these distances are easily compatible with standard lithography techniques. 

In the theory of synchronization the interaction between oscillators and the dynamics of the phase difference, in the simplest cases, can be described by Adler's equation \cite{Adler1946}
\begin{equation}
\frac{\mathrm{d}\psi}{\mathrm{d}t} = \Delta \omega + u\sin \psi \label{eq:commonAdler}
\end{equation}
where $\Delta \omega$ is a difference between frequencies of oscillators and $u$ is proportional to the interaction energy. In the plane of parameters $(\Delta \omega, u)$ the region $-u < \Delta \omega < u$ is the one where Eq. (\ref{eq:commonAdler}) has stable stationary solutions. This zone corresponds to phase locking and frequency entrainment and it is called Arnold tongue \cite{Arnold1991}.

In a second part of this work, we take an interest in capturing a detailed insight on the synchronization process starting from two coupled Thiele equations. These equations describe the vortices motions in their self-induced gyrotropic mode and include spin-transfer term as well as a coupling term \cite{Thiele1973,Khvalkovskiy2009, Gaididei2010, Belanovsky2012}

\begin{align}
G \left( \mathbf{e}_z \times \dot{\mathbf{X}}_{1,2} \right) - k_{1,2}(\mathbf{X}_{1,2}) \mathbf{X}_{1,2} - \mathcal{D}_{1,2} \dot{\mathbf{X}}_{1,2} \nonumber \\ - \mathbf{F}_{STT1,2} - \mathbf{F}_{int}(\mathbf{X}_{1,2}) = \mathbf{0}, \label{eq:Thiele}
\end{align}
here the gyroconstant is given by $G =-2\pi p M_s h/\gamma$, where $p$ is core polarity, $\gamma$ --- gyromagnetic ratio. The confining force is given with $k(\mathbf{X}_{1,2})=\omega_{01,2} G \left(1+a\frac{\mathbf{X}_{1,2}^2}{R_{1,2}^2}\right)$ \cite{Guslienko2006,Ivanov2007} where $R_{1,2}$ are disks radii and the gyrotropic frequency is $\omega_{01,2}=\frac{20}{9}\gamma M_s h/R_{1,2}$. In this study, the Oersted field influence on the dynamics was not taken into account, since at first order it will only shift the self-frequencies of the vortex oscillators and have no influence on the synchronization process. The damping coefficient --- $\mathcal{D}_{1,2} = \alpha \eta_{1,2} G,~\eta_{1,2} = \frac{1}{2}\ln \left( \frac{R_{1,2}}{2l_e} \right)+\frac{3}{8},~\text{where}~l_e=\sqrt{\frac{A}{2\pi M_s^2}}$. The fourth term $\mathbf{F}_{STT1,2}$ is the spin transfer force. For the case of uniform perpendicularly magnetized polarizer $\mathbf{F}_{STT1,2}= \pi \gamma a_J M_s h \left( \mathbf{X}_{1,2} \times \mathbf{e}_z \right) = \varkappa\left( \mathbf{X}_{1,2} \times \mathbf{e}_z \right)$ \cite{Khvalkovskiy2009} where the spin torque coefficient is $a_J=\hbar p_{z} J/(2|e|hM_s)$. The choice was made in this study to apply an equal current density through the two pillars to ensure identical spin transfer forces on the two vortex cores.

\begin{figure}[ht!]
\centerline{\includegraphics[width=8.6cm]{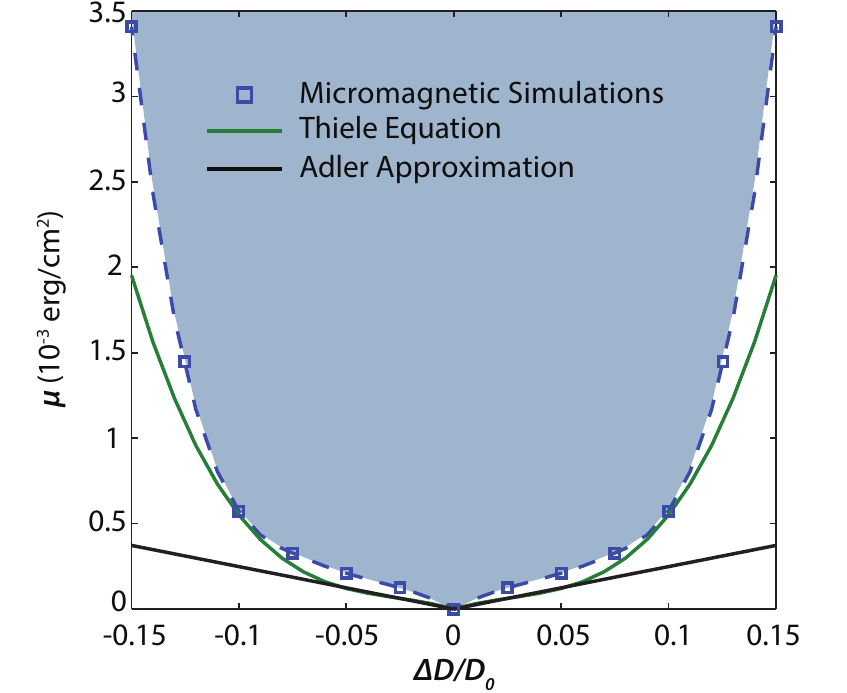}}
\caption{\footnotesize{(Color online) Arnold tongues obtained by micromagnetic simulations (shaded region limited by blue squares), by numerical solutions of Eqs. (\ref{eq:R1})--(\ref{eq:phi2}) (green line), and by Adler approximation (black line)}. The inner part of the tongue represents synchronization region.}
\label{fig:fig3}
\end{figure}

The interpillar interaction is summarized by a dipolar coupling term : the interaction forces between dots are expressed by $\mathbf{F}_{int}(\mathbf{X}_{1,2})=-\mu(L) \mathbf{X}_{2,1}$, where $\mu(L)$ is a coupling parameter depending on the interpillars distance. Here only the first order interactions are considered as developed in a previous paper \cite{Belanovsky2012}. Since small variations of disk diameters do not cause significant changes of dipolar coupling parameter $\mu(L)$, we can fairly estimate it from the results obtained for the case of identical diameters in ref \cite{Belanovsky2012}. The Arnold tongue, synchronization area in the plane ($\Delta D/D_{0},\mu(L)$), extracted with the full micromagnetic simulations is then shown on Fig. \ref{fig:fig3}, as the shaded region limited by dashed line with empty squares.

Using the coupled equations, we now aim at deducing a simple Adler-like equation describing the synchronization process. In polar coordinates $(X_{1,2} \cos \varphi_{1,2}, X_{1,2} \sin \varphi_{1,2})$ Eq. \eqref{eq:Thiele} reads:

\begin{footnotesize}
\begin{subnumcases}{}
\frac{\dot{X}_1}{X_1} = \alpha \eta_1 \dot{\varphi}_1 - \frac{\varkappa}{G} + \frac{\mu}{G} \frac{X_2}{X_1} \sin (\varphi_1-\varphi_2) \label{eq:R1}\\
\dot{\varphi}_1 = -\frac{k(X_1)}{G} - \alpha \eta_1 \frac{\dot{X}_1}{X_1} - \frac{\mu}{G} \frac{X_2}{X_1} \cos (\varphi_1-\varphi_2) \label{eq:phi1}\\
\frac{\dot{X}_2}{X_2} = \alpha \eta_2 \dot{\varphi}_2 - \frac{\varkappa}{G} - \frac{\mu}{G} \frac{X_1}{X_2} \sin (\varphi_1-\varphi_2) \label{eq:R2}\\
\dot{\varphi}_2 = -\frac{k(X_2)}{G} - \alpha \eta_2 \frac{\dot{X}_2}{X_2} - \frac{\mu}{G} \frac{X_1}{X_2} \cos (\varphi_1-\varphi_2) \label{eq:phi2}
\end{subnumcases}
\end{footnotesize}

Let us make some simplification for the small diameters differences $(\Delta D / D_0 < 0.05)$:
a) The condition $\Delta D / D_0 \ll 1$ gives us possibility to linearise the following terms as
\begin{footnotesize}
\begin{align}
& k_{1,2}(X_{1,2}) \approx G\left[\omega_0\left(1 \mp \frac{1}{2} \frac{\Delta D}{D_0}\right) + \omega_0 a \frac{X_{1,2}^2}{R_0^2} \left(1 \mp \frac{3}{2}\frac{\Delta D}{D_0}\right)\right]\nonumber \\
& \eta_{1,2} \approx \eta_0 \pm \frac{1}{4}\frac{\Delta D}{D_0},\nonumber
\end{align}
\end{footnotesize}
where $\omega_0$ is gyrotropic frequency for disk with radius $R_0 = 100$ nm and the damping $\eta_0 = \frac{1}{2}\ln \left( \frac{R_0}{2l_e} \right)+\frac{3}{8}$.
b) Assuming that the steady state vortex radii almost doesn't differ from its mean value, thus one can write $X_{1,2} = X_0(1 \pm \varepsilon)$, where $X_0 = (X_1 + X_2)/2 = 60$ nm (this value was obtained by micromagnetic simulations for the case of same diameters with $D_0 = 200$ nm), and $\varepsilon = (X_1-X_2)/(X_1+X_2) \ll 1$ at the limit cycle. This assumption was confirmed by micromagnetic modeling for small diameters differences.
c) The synchronization of STVOs with small diameters difference appears on large interpillar distances, on these distances the interaction between disks is quite weak and we can say that $\mu \ll G\omega_0 a r_0^2$, where $r_0 = X_0/R_0$ [our calculations showed that for $L>400$ nm $\mu \sim 10^{-4}$ erg/cm$^2$ and $G \omega_0 a r_0^2 \sim 0.1$ erg/cm$^2$].

These approximations allow us to transform Eqs. \eqref{eq:R1}-\eqref{eq:phi2} to the second order differential equation for phase difference $\psi = \varphi_1 - \varphi_2$:
\begin{equation}
\frac{1}{2a\omega_0 r_0^2} \ddot{\psi} + \alpha \eta_0 \dot{\psi} + 2 \frac{\mu}{G} \sin \psi = \frac{1}{2}\alpha \frac{\Delta D}{D_0} \omega_0 (1 + ar_0^2). \label{eq:psi}
\end{equation}

At the limit cycle the second derivative of $\psi$ in Eq. \eqref{eq:psi} tends to zero. The equation of motion then reads

\begin{equation}
\frac{\mathrm{d} \psi}{\mathrm{d} t} =  \frac{1+ar_0^2}{2\eta_0}\frac{\Delta D}{D_0}\omega_0 - \frac{2\mu}{G\alpha \eta_0} \sin \psi. \label{eq:adler2}
\end{equation}
Eq. (\ref{eq:adler2}) has the stationary solutions when
\begin{equation}
|\mu| \leq G\frac{\alpha}{4}(1+ar_0^2)\frac{\Delta D}{D_0}\omega_0 \nonumber
\end{equation}

The synchronization region for Eq. \eqref{eq:adler2} is added on Fig. \ref{fig:fig3} (black line). It is seen that when $\Delta D / D_0 \geq 5\%$ the solution of Thiele equations differs from micromagnetic simulations  even qualitatively. Therefore, our linearisation cannot be extended to the cases of significant diameter mismatch. To complete our study and capture such strongly asymmetric regimes, we finally evaluate the validity of the coupled Thiele equations, without the latter approximations, to describe the synchronization dynamics.

The Eqs. \eqref{eq:R1}, \eqref{eq:phi1}, \eqref{eq:R2} and \eqref{eq:phi2} can be solved numerically using coupling parameter $\mu$ derived from micromagnetic simulations \cite{Belanovsky2012} and the corresponding phase-diagram can be extracted (green line in Fig. \ref{fig:fig3}). This numerical boundary is added as a green line in Fig. \ref{fig:fig3}). As can be seen, Eqs. \eqref{eq:R1}-\eqref{eq:phi2} give us reliable results for diameters difference bigger than $5 \%$, further than with the simple Adler-like approach.

With the further increase of diameters difference the mismatch between the results obtained using Eqs. \eqref{eq:R1}-\eqref{eq:phi2} and the ones derived from micromagnetic simulations becomes more significant. These differences come from the fact that the Thiele equation was initially developed to describe steady oscillations regime, with unperturbed orbit radius \cite{Ivanov2010,Gaididei2010}. Therefore our equations cannot perfectly describe the core dynamics when perturbed by dipolar interaction, especially when the self-frequency difference between oscillators is strong. However this approach can describe reliably the synchronization process until the diameters differences of about $12\%$ that are much more than standard error of a state-of-the-are fabrication process.

In summary, we have shown the possibility to synchronize two STVOs with a frequency difference due to pillars diameters difference. Our micromagnetic simulations have shown that the phase locking of this system appears at the interpillar distances below a critical one $L_{cr}$ which depends on diameter difference $\Delta D / D_0$. We have obtained the phase diagram (Arnold tongue) which demonstrates the transition between synchronized and unsynchronized regions. Also we have provided a quantitative analytical analysis based on Thiele equations. Our study has shown that the system becomes strongly non-Adlerian once the diameters difference increases significantly. However, for $\Delta D / D_0 \leq 5\%$, the synchronization phase diagram can be described qualitatively using an Adler-like equation (\ref{eq:adler2}). Although further increase of diameters difference drives the synchronization dynamics into non-Adlerian regime, it can still be well described by numerical integration of non-linearized Thiele equations. Only at a huge diameters differences (more than $12\%$), which is far beyond the maximum acceptable error in the state-of-the-art fabrication process, Thiele-based approach fails to describe the synchronization process. The reason of such increasing complexity of synchronization regimes  with the increase of the system's asymmetry might be the excitation of strongly nonlinear modes.

The authors thank Alexey V. Khvalkovskiy and Lee C. Phillips for helpful discussions.
The work is supported by RFBR Grants No. 10-02-01162 and
No. 11-02-91067, CNRS PICS Russie No. 5743 2011,
Dynasty Foundation, and the ANR agency (VOICE
PNANO-09-P231-36). F.A.A. acknowledges the Research
Science Foundation of Belgium (FRS-FNRS) for financial
support (FRIA grant).

\end{document}